%
%
\input epsf

%

\rightline {RU--91--31}
\rightline{FSU-SCRI-91-94}
\vskip 1.5cm
\centerline{\bf HOW TO PUT A HEAVIER HIGGS ON THE LATTICE$^{\S}$}
\vfootnote\S{Submitted to Physical Review Letters}
\vskip 2.5cm
\centerline {by}
\vskip 1cm
\centerline{Urs M. Heller${}^{\dag}$, Herbert Neuberger${}^{\ddag}$, and
 Pavlos  Vranas${}^{\dag }$}
\vskip 1cm
\centerline{\it ${}^{\dag}$ Supercomputer Computations Research Institute}
\centerline{\it The Florida State University}
\centerline{\it Tallahassee, FL 32306}
\vskip .2cm
\centerline{\it ${}^{\ddag}$ Department of Physics and Astronomy}
\centerline{\it Rutgers University}
\centerline{\it Piscataway, NJ 08855--0849}
\vfil
\centerline{\bf Abstract}
\bigskip
Lattice work, exploring the Higgs mass triviality bound,
seems to indicate that a strongly interacting scalar sector in the
minimal standard model cannot exist while low energy
QCD phenomenology seems to indicate that it could. We
attack this puzzle using the $1/N$ expansion and discover a simple criterion
for selecting a lattice action that is more likely to produce a heavy Higgs
particle. Our large $N$ calculation suggests that the Higgs mass bound
might be around $850~GeV$, which is about 30\% higher than previously obtained.
\eject

The recent quantitative results that emerged from the investigation of
triviality in the $\lambda (\vec\phi ^2)^2$  theory on the lattice [1] have
given an upper bound on the ratio of the Higgs mass, $m_H$, to the weak
scale, $f_\pi = 246 ~GeV$, of about 2.5--2.7. In the hypothetical world of
QCD with just the up-down doublet of quarks with zero masses, which is
described by the same effective theory at low energies, one believes, on
phenomenological grounds, that the above ratio would be significantly
larger, equal to about 6--8. This is admittedly an extreme situation
because the sigma enhancement in $\pi$ $\pi$ scattering doesn't quite
qualify to be called a resonance, and, it comes accompanied by a host of
other resonances, representing all kinds of nontrivial ``cutoff effects''.
Still, there is a large gap between 2.7 and 6 and we are left with
essentially two alternatives: either the QCD example provides a totally
unreasonable estimation for the bound, or, we admit that the lattice
result, while solidly established within a subclass of actions, is
relatively far from a realistic estimate for the bound. We would have much
less cause for worry if the lattice bound for the ratio came out to be
about 4 (this would correspond to $m_H \simeq 1~ TeV$) or more [2]. 

This issue could be addressed in principle by an extensive search of all
possible lattice actions. In practice one needs some theoretical indication
as to what kind of regularized action is likely to produce a higher bound.
If the physical picture behind the particular form of the suggested action
is reasonable, an investigation of its consequences would help resolve the
above puzzle by either refuting the existing lattice bound or by
strengthening the contention that the QCD example is grossly exaggerated. 

The purpose of the present note is to make a proposal for such a
generalized action and sketch the theoretical basis for it. The basic point
of the proposal is that one should put on the lattice nonlinear $O(4)$
models that have a naive continuum limit with freely adjustable four
derivative couplings and study them nonperturbatively. We solve the model
in the large $N$ limit of $O(N)$ and find the region in coupling space
where the highest bound is expected; we test the extent to which the
results are dependent on the regularization and extract the basic features
that are generic. The results can be explained by a rough physical picture
as follows: When the regularized model is nonlinear one has to think about
the Higgs resonance as a loosely bound state of two pions in an $I=0$,
$J=0$ state. Pions in such a state attract because superposing the field
configurations corresponding to individual pions makes the state look more
like the vacuum and hence lowers the energy. The four derivative terms in
the action can add to this attraction; if the attraction is increased to a
certain level the bound state will become massless and stable. This occurs
while the system is still relatively deep in the broken phase; hence the
weak scale $f_\pi$ is nonvanishing. The appearance of this massless bound
state signals a new kind of phase transition; the ordinary phase transition
occurs when both $f_\pi$ and $m_H$ vanish and leads to a scale invariant
theory at the transition - here we still have a scale and therefore scale
invariance seems to be spontaneously broken, with the massless boundstate,
the Higgs particle, playing the role of the expected dilaton. This
phenomenon is reminiscent of the results obtained in three dimensions by
Bardeen, Moshe and Bander [3]. As far as the Higgs mass bound goes, it is
clear that we have to try to be as far away as possible from this new
transition point; we wish to induce enough repulsion between the pions to
delay the creation of the bound state as much as possible. Of course, the
interaction is still attractive overall, because only the order $p^4$ term
in the interaction can be tampered with; the order $p^2$ and $p^4 \log
(p^2)$ terms are fixed by current algebra and are attractive [4]. Moreover,
in the critical regime the model still is governed by the $\lambda ({\vec
\phi} ^2)^2 $ trivial fixed point and is expected therefore to generate a
one particle state corresponding to the $\sigma$ field component. While it
is quite possible that the above described scale invariance breaking
transition cannot actually materialize at $N=4$ (we shall see that at
$N=\infty$ it does appear in all regularizations that we tested), the basic
trend for the magnitude of the triviality bound must be the same. 

Assuming global $O(4)$ symmetry and that the leading cutoff effects are
represented by dimension six operators, one arrives, after eliminating
couplings that won't lead to observable effects in the S-matrix to leading
order in the inverse cutoff, to a space of actions that depends on 4
parameters [5]. One may try to induce the dependence on the associated
renormalized operators by simply writing identical forms for terms in the
bare action. However, we are looking for a bound on the physical four point
coupling and it is likely that it obtains in the limit that the bare four
point coupling is taken to infinity. When the bare coupling becomes
infinite the bare model is reduced to a nonlinear model; this has the
effect that the naive forms of the dimension six operators in terms of bare
linear fields become trivial. The parameterization in terms of linear
fields and operators of naive dimension six breaks down completely at the
point where the bare theory becomes nonlinear (we have tested this
explicitly in the linear model, within $1/N$, and with a particular form of
a bare dimension-six term in the Lagrangian). So we were expecting to be
left with three freely adjustable parameters (within some range) but we
ended up with only one. This is the case that has been investigated
nonperturbatively until now [1]. To include the effects of dimension six
operators at physical scales we need to put in operators of higher order in
derivatives in the bare, nonlinear, theory that describes the physics at
cutoff scales. There are exactly two additional terms that one can write
down if one restricts the number of derivatives to four [4]; thus the right
number of  parameters is obtained. This time, unlike in the linear case,
the effect of the bare couplings has no reason to disappear. 

It is well known that the $1/N$ expansion, at least to leading order,
doesn't have to work well quantitatively simply because there is a factor
of $N+8$ in the coefficients of the $\beta$ function and therefore $N=4$ is
badly approximated by $N=\infty$ [6-8]. We shall say more about this later;
for the moment keep in mind that we are interested mainly in relative
effects, that is, in how the bound varies with the action, and less in the
absolute magnitude of the bound. To get the latter one would have to resort
to numerical means.

If a strongly interacting scalar sector could be produced on the lattice
its investigation would require some nontrivial extension of the tools that
were used until now. However, if we only wish to establish the existence of
a strongly interacting regime, we can do that indirectly, while restricting
our attention just to weak couplings. For example, a strongly coupled Higgs
sector was excluded on lattices with a pure nearest-neighbor coupling, by
showing that, for a ratio $m_H /f_\pi$ of only 2.5, one already had $m_H /
\Lambda $ equal to 0.5, leaving no ``room'' below the cutoff $\Lambda$ for
further growth in the coupling; if we could show that a different action
would, at the same value of $m_H /f_\pi$, give $m_H / \Lambda $ equal to
1/10 say, our case would be made. Therefore, our subsequent study is
restricted to the weakly coupled sector. Generically, we expect there that 
$$
{m_H \over \Lambda } \approx C(b) \exp [ - 16\pi^2 f_{\pi}^2 /(Nm_H^2 ) ],
\eqno(1)
$$
where $C(b)$ is a function of the additional parameters we introduced in
the action. We aim to show that by varying $b$ $C(b)$ can be significantly
lowered below the value it had when no four derivative terms were included
(this is the only case that has been investigated to date). Our finding is
best summarized in a qualitative self-explanatory diagram that we checked
for a class of Pauli-Villars regularizations and for the $F_4$ lattice
regularization (the generalization of the calculations to the hypercubic
lattice should be a simple exercise). 

\midinsert
\epsfxsize=4in
\centerline{\epsffile{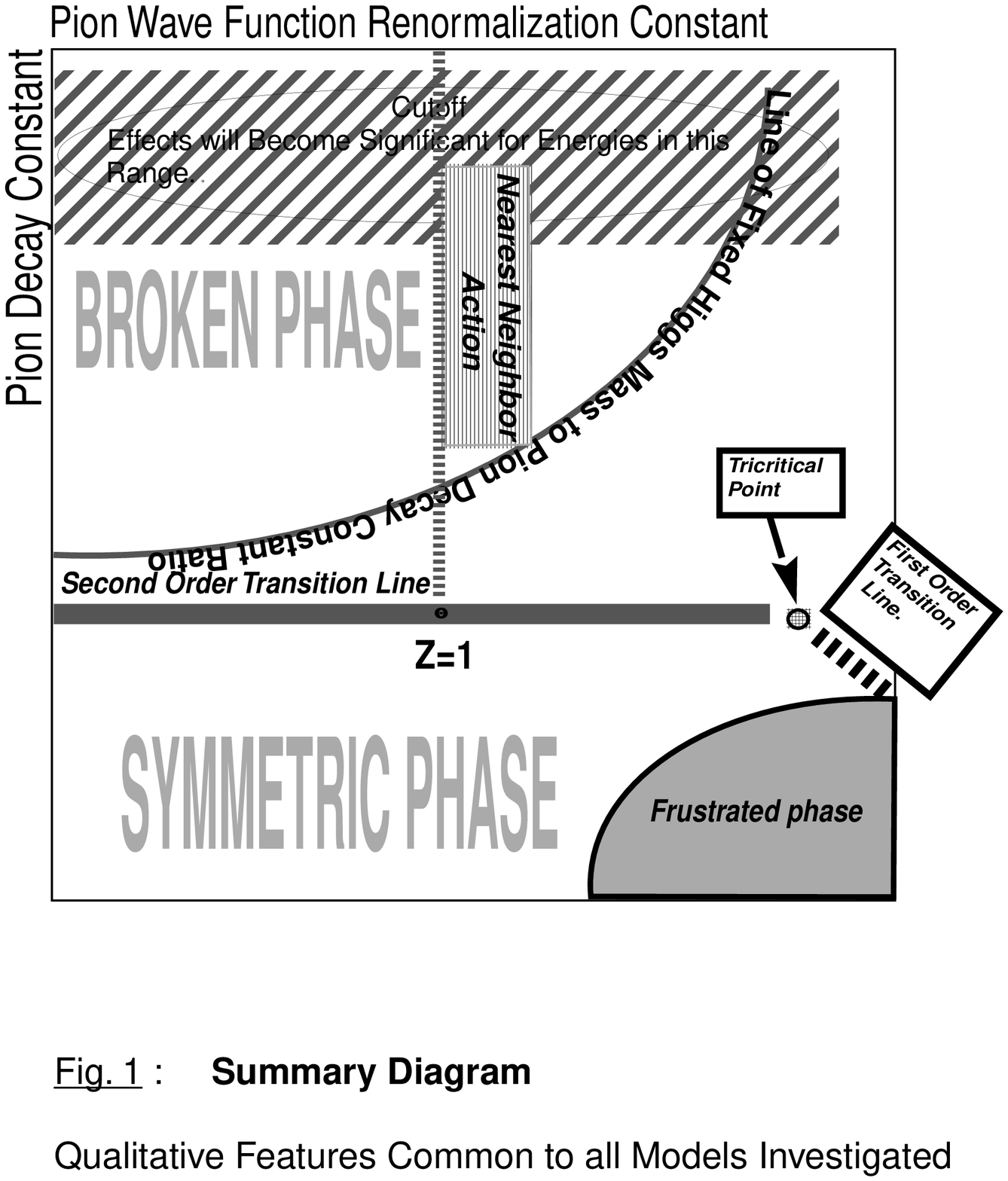}}
\endinsert

We first look at the models with Pauli-Villars, or higher derivative,
regularization. We choose to replace the ordinary propagator by $1/[p^2(1+
(p^2 /M^2 )^{n} )]$ where $M$ is a ``bare'' cutoff (as opposed to $\Lambda$
the physical cutoff to be defined below) and $n \ge 3 $ is an integer. The
limit $n=\infty$ corresponds to a sharp momentum cutoff and is expected to
be singular. For any finite $n$ the bare propagator has $n$ ``ghost'' poles
all situated on the unit circle in the complex $s/M^2$ plane. We shall
define the physical cutoff, $\Lambda$, as the square root of the modulus of
the location of the closest complex pole in $s$ in the full pion propagator
(at leading order in $1/N$). The action, written in Euclidean space, is: 
$$
S=\int_x ~~ \left [{1\over 2} \vec \phi ~g(-\partial^2 )\vec \phi -
{b_1 \over {2N}} (\partial_\mu
\vec \phi \cdot \partial_\mu \vec \phi )^2 -
{b_2 \over {2N}} (\partial_\mu
\vec \phi \cdot \partial_\nu \vec \phi - {1\over 4} \delta_{\mu , \nu }
\partial_\sigma \vec \phi \cdot \partial_\sigma \vec \phi )^2 \right]
\eqno(2)
 $$
where $g(-\partial^2 )= (-\partial^2 )[1+ (-\partial^2 / M^2 )^n ] $, and
the field is subjected to the constraint ${\vec \phi} ^2 = N\beta $. The
partition function is ${\cal Z} = \int [d {\vec \phi} ] \exp [-S]$. 

Using the technique employed in ref. [9]    we find the phase diagram of
the model paying particular attention to the competition between various
saddles. We set 
$$
\vec \phi = {\sqrt {N}} {\vec v} + {\hat v} H(x) + {\vec \pi}(x)
\eqno(3)
$$
with ${\hat v} = {\vec v} / v,~ v=|{\vec v}|,~{\hat v} \cdot {\vec \pi}
=0,~~ \int_x H(x) =0,~~\int_x {\vec \pi }(x) =0 $. We define the pion wave
function renormalization constant $Z_{\pi}$  for the field $\vec \pi$ as
usual. The pion decay constant $f_\pi$ can be easily gotten from $v$ and
$Z_\pi$. 

 It turns out that it is more convenient, for the broken phase, to
parameterize the theory by $Z_\pi$: 
$$
b_1= \left( {{4\pi}\over{M^2}} \right)^2 {{n\sin  (2\pi /n)}\over {2\pi}}
{{Z_\pi -1}\over
{Z_{\pi}^{2(n-1)/n}}}
\eqno(4)
$$
where $0 \le Z_\pi \le 2(n-1)/(n-2)$. In this range, $b_1$ increases
monotonically with $Z_\pi$ and at the larger end-point we have $\partial
b_1 / \partial Z_\pi =0$. The portion of the phase diagram we are
interested in does not depend on $b_2$ because giving an expectation value
to the $O(N)$ singlets associated with $b_2$ would break Lorentz
invariance. The other coupling is best replaced by the unitless weak scale,
$F=f_\pi / \Lambda $. The physical cutoff, $\Lambda$, is given by $\Lambda
= M/Z_\pi ^ {1/(2n)}$. From the large $N$ saddle point equations we obtain
the expectation value $v$ which is related to $f_\pi$ by $ v^2 =Z_\pi
f_{\pi}^2$ resulting in: 
$$
{{ (4\pi F )^2}\over{N}} = \left( {{4\pi}\over{M}} \right)^2 \beta {Z_{\pi}}
^{(1-n)/n} - {{\pi}
\over {n\sin (\pi /n)}} .
\eqno(5)
$$

There is a second order transition line at $F=0$ and, on it, at the end of
the allowed range of $Z_\pi$, there is a tricritical point where the second
order line turns into first order. At the tricritical point one obtains a
massless pole in the two point correlation function of the trace of the
energy-momentum tensor. We shall not be concerned with the region beyond
the tricritical point. The calculation of $m_H$ proceeds in the standard
manner; it turns out that one has to find the complex zeros of the
determinant of a four by four matrix. Since we restrict our attention to
the perturbative regime, i.e. $F, p << 1$ we can replace the quantity $m_H$
by the technically more accessible real quantity $m_R$, defined from the
smallest positive zero in $s=-p^2$ of the real part of the determinant by
$m_R =\sqrt s$. We are really interested in $M_R =m_R /\Lambda $. If we
also take $M_R <<1$ of the same order (up to logs) as $F$, the fourth row
and column of the matrix decouple and the dependence on $b_2$ disappears.
We need the expression for the ``bubble'' diagram 
$$\eqalign{
B_{PV} (p) &= \int {{d^4 k}\over {(2\pi )^4}} {1\over{g(k-p/2) g(k+p/2)}}\cr 
& =
-{1\over{16 \pi^2}} \left[ \log (p^2 / M^2 ) -{{n-1}\over n} \right]
+ O (p^2 / M^2 )\cr}
\eqno(6)
$$
which holds for any $n < \infty$ (for $n=\infty$ the correction term goes
as $\sqrt { p^2 / M^2 }$ and represents a non-local term in the effective
action as expected when employing a sharp momentum cutoff). One has then: 
$$
M_R^2 \approx \exp \left[ -{2\over N} \left( {{4\pi F}\over{M_R}} \right)^2
\right] ~ C^2 (Z_\pi ) , 
$$
$$
C(Z_\pi ) = \exp \left[ {{n-1}\over{2n}} \left( 1- {\pi\over{n \tan (\pi
/n)}} {{1-Z_\pi } \over {1- Z_\pi (n-2)/(2(n-1)) }} \right) \right] . 
\eqno(7) 
$$
There is no four derivative term when $Z_\pi =1$; the minimal $C$ obtains
at $Z_\pi =0 $ which is as far away from the tricritical point as possible:
$$
{{C(Z_\pi =1)}\over{C(Z_\pi =0 )}} = \exp
\left[ \left( {{n-1}\over{2n}} \right) {\pi\over{n\tan{(\pi/
 n)}}} \right] ~ \le ~ \exp (0.5) \approx 1.65 .
 \eqno(8)
$$

We have obtained some increase in the cutoff but not a factor of about 5 as
we would have liked. However, we know from the work of Einhorn [7] that
even at $b_1=0$ this particular large $N$ model will have a strongly
interacting sector; therefore it may be that there is no room left to
dramatically increase the cutoff even further. It is also clear that if it
turned out that the lattice actions that were so far investigated really
correspond to an action with $b_1 >0 $ in Pauli-Villars regularization, a
much larger variation could be obtained because $C$ diverges at some
finite, positive $b_1$. $~ b_1 >0$ means that the extra term has induced
more attraction in the $I,J=0$ channel and the divergence in $C$ occurs at
the tricritical point where the resonant bound state becomes massless and
stable. 

We turn now to the lattice. Since our objective here is to provide a useful
indication for what kind of actions would be instructive to study by
nonperturbative means, we pick the lattice regularization that is best
suited for this purpose, namely we put the theory on an $F_4$ lattice. This
lattice has a larger symmetry group than the more commonly used hypercubic
lattice which ensures that no Lorentz invariance breaking four derivative
terms appear in the (naive) continuum limit of its lattice actions. The
$F_4$ lattice can be viewed as a hypercubic lattice (${\rm Z}^4$) with unit
lattice spacing from which all the sites with an odd sum of coordinates
have been erased [5]. We shall denote sites on $F_4$ by $x, x',x''$, and
links by $<x,x'>,l,l'$. The action is chosen as 
$$\eqalign{
S=& -2N\beta_0 \sum_{<x,x'>} \vec \Phi (x) \cdot \vec \Phi (x') -N\beta_1
\sum_{<x,x'>} \left[
\vec \Phi (x) \cdot \vec \Phi (x')  \right]^2 \cr
&\quad\qquad -N{{\beta_2}\over{48}} \sum_x
\left[\sum_{{l\cap x \ne
\emptyset}\atop {l=<x,x'>} } \vec \Phi (x) \cdot \vec \Phi (x') 
\right]^2 .\cr} 
\eqno(9)
$$
Here the field is constrained by ${\vec \Phi}^2 (x) =1$. To express the
action in a form closer to the one we employed before we rescale the field
$ \vec \phi = \sqrt {6N(\beta_0 +\beta_1 +\beta_2)} \vec \Phi $ (we only
consider the region $\beta_0 +\beta_1 +\beta_2 > 0$). The action has the
same form as above (eq. (2)), up to terms that contain fourth order
derivatives, if one identifies: 
$$\eqalign{\openup2\jot
g(k_\mu) =& {1\over 6} \sum_{\mu \ne \nu } 
\left[2-\cos (k_\mu +k_\nu ) -\cos (k_\mu
-k_\nu )\right] \approx k^2 -{{(k^2)^2 }\over 12} + O(k_\mu^6) ,\cr
b_1 =& {1\over{48}} {{\beta_1 +\beta_2 }\over {(\beta_0 +\beta_1 +\beta_2 )^2}}
 ,~~~~~~~~~~~~~~~~~
b_2 = {1\over{36}} {{\beta_1}\over {(\beta_0 +\beta_1 +\beta_2 )^2}} .\cr}
\eqno(10)
$$

Again $b_2$ plays no role and one may set $\beta_1$ to $0$. The situation
is essentially the same as in the Pauli-Villars case. $b_1$ can be traded
for the pion wave function renormalization constant, 
$$
b_1 ={1\over{Z_\pi}}-{1\over{Z_\pi^2}},
\eqno(11)
$$
where $Z_\pi$ varies between $0$ and $2$. The standard non-linear action
corresponds again to $Z_{\pi} = 1$ and, on the line where the pion decay
constant $F$ vanishes, $Z_{\pi} =2$ corresponds to the tricritical point.
$F$ is given by: 
$$
{{F^2}\over N} = {{6(\beta_0 + \beta_1 +\beta_2)}\over {Z_\pi}} - r_0 ,
~~~~~~~~~~~ r_0 = \int_k {1\over{g(k)}} = 0.13823 .
\eqno(12)
$$
The momentum integral is over the $F_4$ Brillouin zone and a factor of
$1/(2\pi) ^4$ has been absorbed in the definition of $\int_k $; the
numerical value is taken from ref. [5]. 

We proceed now to find $M_R$ defined exactly as before; for this we need
the lattice ``bubble'' which can be easily evaluated using the results of
ref. [5] (but note a missing minus sign in the value of $r_1$ in equation
(4.12) there): 
$$\eqalign{
B_{F_4} (p) =&  -{1\over{16 \pi^2}}  \log (p^2)   +c_1 +c_2 p^2 +
O (p^4 \log p^2 ),\cr
&\cr
c_1=&~~ 0.0466316,~~~~~~~~~~~~~~~~~~~~~~~~~~~~~~~~~~~c_2=0.0005497 .\cr}
\eqno(13)
$$
Unlike in the Pauli-Villars case, the cutoff (defined for example by a
certain amount of violation of Euclidean rotational invariance in the full
pion propagator) is independent of the couplings in the action. 

This time we shall be interested to go beyond the $O(1)$ term and that's
why we evaluated the next correction in $p$ to the bubble (13). Ignoring
this correction for the moment we get the usual formula for $M_R$ (eq.
(7)), but the coefficient $C$ is now given by, 
$$
C(Z_\pi ) = \exp \left[ 8\pi^2 \left( c_1 - r_0^2 {{1-Z_\pi} \over
{1-Z_\pi /2 }} \right) \right],
\eqno(14)
$$
leading to:
$$
{{C(Z_\pi =1)}\over{C(Z_\pi =0 )}} = \exp ( 8\pi^2 r_0^2 )
 \approx 4.521.
 \eqno(15)
$$
We see that we have obtained a much larger variation.

In practice, for vectorization purposes, it would be better to have an
action that involves only nearest neighbors on the $F_4$ lattice. Nothing
in what we have said until now changes if we replace the $\beta_2$ term in
the action by 
$$
N~{{3\beta_2}\over{8}} \sum_x
\sum_{{<ll'>}\atop {l,l' \cap  x \ne
\emptyset ,~ l\cap x' \ne\emptyset ,~ l'\cap x'' \ne\emptyset ,~ x,x',x''
~{\rm all~ n.n.}} } \left[ \left( \vec \Phi (x) \cdot \vec \Phi (x') \right)
\left( \vec \Phi (x) \cdot \vec \Phi (x'') \right) \right] .
\eqno(16)
$$

To check sensitivity to higher order terms in $F^2$ and/or $M_R^2$, while
still neglecting $b_2$, we go to the next higher order in $F, p, M_R $ when
we find the Higgs mass. It turns out that the numerical correction is small
(of the order of 1--2 \% ) even for $M_R \approx 1$. For the usually
studied action with $b_1$ and $b_2 = 0$ we can check how well the $1/N$
expansion works, on the quantitative level, in the region of main interest,
namely $m_R / f_\pi  \approx 2.0 -2.5 $; in this region we have numerical
data at $N=4$ (see last reference in [1]). We find agreement to within 25
percent for the $M_R /F$ ratio as a function of $M_R$; the large $N$
results are systematically larger than the $N=4$ ones but the relative
change in the ratio $M_R /F$ when $M_R$ is varied is more accurately
predicted. This is enough to convince us that our large $N$ results ought
to be taken seriously as an indication for what is going to happen in the
real system. For example, we would predict that if we limit cutoff effects
by the reasonable [5] requirement $M_R \le \sqrt{2} /2$ we find, with
$N=4$: 
$$
{{M_R}\over{F}} \le {{2\pi}\over {\sqrt{\log \left[\sqrt{2} C(Z_\pi )
\right]}}} .
\eqno(17)
$$
This gives for the nearest neighbor case ($Z_\pi =1$) $M_R /F$ =3.1 while
for the extreme case ($Z_\pi =0$) one finds $M_R /F =4.0$. For $N=4$ the
numerical result is smaller by about 20\%. We therefore are led to
conjecture that for the simplest case with $\beta_1 =0$ the bound would
increase by 30\% to about $850~ GeV$. 

We should comment here on the possible relationship between this note and
the work based on effective Lagrangians that was carried out in the context
of the investigation of the scalar sector of the standard model by several
groups recently [10]. Although in both cases one employs a nonlinear
Lagrangian of identical forms (at least for $N=4$) there is a fundamental
difference that ought to be stressed: In the effective Lagrangian approach
one uses {\it renormalized\/} perturbation theory as defined by Weinberg
[4] to summarize all relations between soft pion amplitudes to any finite
order in the external momenta of the pions; one then tries to extrapolate
[10] these results by several schemes, all attempting to approximately
impose unitarity and crossing, to the region of energies dominated by the
tails of the lowest resonances. At no point does this approach consider the
question of whether the physics so defined can be consistently extended to
the cutoff (see second reference in [4] for some discussion). Technically
speaking, chiral renormalized perturbation theory, very much like ordinary
perturbation theory, does not provide the relation between the parameters
in the bare Lagrangian (i.e. cutoff scale physics) and the parameters that
describe low energy properties; it only provides a very large number of
constraints between all the low energy physical processes. Thus, questions
having to do with triviality are completely outside the reach of these
approaches. In more intuitive terms, we are taking here very seriously a
model defined by a chiral Lagrangian, all the way to the highest energies,
and seek for constraints on the allowable low energy effects that result
from the requirement that they occur below the cutoff. If the cutoff is
very high relative to the lowest resonance we are bound, by the general
principles of the renormalization group, to view this as just another
regularization of $\lambda ({\vec \phi}^2 )^2 $ and hence a perfectly
reasonable framework for investigating consequences of triviality. 

We have not discussed in any detail the term with coupling $b_2$. We have
seen that it has no effect in the perturbative regime; however, when terms
beyond the equivalent of the $O(p^2)$ correction to the bubble are taken
into account, $b_2$ comes into play and can become important. The structure
of the term indicates that $b_2$ is the right dial to tune if we wish to
make our theory look more like QCD. This suggests that it might help to
make the Higgs mass bound even higher. But in the present note we were
interested mainly in estimating the Higgs-mass triviality bound when
nothing special (technicolor-like) has been imposed on the theory. Further
discussion of the $b_2$ term would take us too far afield. We shall also
omit any discussion of subleading terms in the $1/N$ expansion;
particularly interesting would be the corrections to $Z_\pi$ and to the
phase structure in the vicinity of the tricritical point. These topics and
a comprehensive analysis of Renormalization Group flows in the models under
consideration will hopefully be addressed in future work. 

We should also comment here on the fact that it is well known that once the
scalar sector is strongly interacting, looking at $m_R$ rather than $m_H$,
is a bad approximation due to the large width [11]. This effect leads to a
saturation of the growth of $m_H$ much before $m_R$ stops growing and can
be best seen in the $1/N$ expansion, as shown by Einhorn [7]. It is
important to realize that this sort of bound has nothing to do with
triviality; it appears even when the issue of triviality is totally
ignored, for example, when the bare four point coupling in the linear model
is allowed to become negative and the theory has a tachyonic instability.
Our calculations can all be done for a relatively narrow Higgs, and all
they aim to show is that a strongly interacting scalar sector in the
minimal standard model has not yet been convincingly ruled out. The
relation between $m_R$ and $m_H$ alluded to above may very well hold in
such a sector, if only we could show that one exists. 

In summary, we propose that the scalar sector of the minimal Standard Model
can be made to have a stronger renormalized self coupling by introducing,
at the cutoff level, four-derivative couplings that induce some repulsion
between Goldstone Bosons. Within the $1/N$ expansion this increases the
Higgs mass bound, compared to the usual lattice result, by up to about
30\%. Obviously it would be extremely interesting to study, by numerical
means, whether this conclusion persists for the physically relevant case of
$N = 4$. This would help resolve the puzzle presented at the beginning of
this Letter. We hope to address this question in the future. 

\bigskip
\noindent {\bf ACKNOWLEDGEMENTS.} 

This research was supported in part by
the Florida State University Supercomputer Computations Research Institute
which is partially funded by the U.S. Department of Energy through DOE
grant \# DE-FG05-85ER250000 (UMH and PV) and under DOE
grant \# DE-FG05-90ER40559 (HN). UMH also acknowledges partial support by
the NSF under grant \# INT-8922411 and he would like to thank F. Karsch and
the other members of the Fakult\"at f\"ur Physik at the University of
Bielefeld for the kind hospitality while part of this work was done. HN
would like to thank F. David and M. Einhorn for useful discussions. 

\vfil\eject
\noindent{\bf REFERENCES}
\smallskip
\parskip=8pt

\item {[1]} M. L\"uscher and P. Weisz, {\it Phys. Lett.},  {\bf B212} 
(1988) 472;
J. Kuti, L. Lin and Y. Shen, {\it Phys. Rev. Lett.}, {\bf 61} (1988) 678; A.
Hasenfratz, K. Jansen, J. Jers\'ak, C. B. Lang, T. Neuhaus, H. Yoneyama,
{\it Nucl. Phys.},  {\bf B317} (1989) 81; G. Bhanot, K. Bitar, {\it Phys. Rev.
Lett.}, {\bf 61} (1988) 427; G. Bhanot, K. Bitar, U. M. Heller, H. Neuberger, {\it
Nucl. Phys.}, {\bf B353} (1991) 551.

\item {[2]} see the {\it The Standard Model Higgs Boson} edited by M. B.
Einhorn, {\it CPSC} Vol. {\bf 8}, Elsevier Pub. 

\item {[3]} W. Bardeen, M. Moshe, M. Bander, {\it Phys. Rev. Lett.}, {\bf 52 }
(1983) 1188. 

\item {[4]} S. Weinberg, {\it Physica}, {\bf 96A} (1979) 327; H. Georgi, {\it Weak
Interactions and Modern Particle Theory} (1984) Addison-Wesley Pub.

\item {[5]} G. Bhanot, K. Bitar, U. M. Heller, H. Neuberger, {\it Nucl.
Phys.}, {\bf B343}, (1990) 467. 

\item {[6]} R. Dashen, H. Neuberger, {\it Phys. Rev. Lett.}, {\bf 50} (1983) 1897.

\item {[7]} M. Einhorn, {\it Nucl. Phys.}, {\bf B246} (1984) 75. 

\item {[8]} L. Lin, J. Kuti, Y. Shen, in {\it Lattice Higgs Workshop, FSU,
May 16--18 1988}, Editors B. Berg., et. al. 

\item {[9]}  F. David, D. Kessler, H. Neuberger, {\it Nucl. Phys.}, {\bf B257}
[FS14] (1985) 695. 

\item {[10]} J. Bagger, S. Dawson, G. Valencia, BNL -- 45782,
Brookhaven Nat. Lab. preprint; S. Dawson, G. Valencia, {\it Nucl. Phys.},
{\bf B352 } (1991) 27; J. F. Donoghue, C. Ramirez, G. Valencia, {\it Phys. 
Rev.}, {\bf D39} (1989) 1947; A. Dobado, M. Herrero, J. Terron, 
CERN--TH.5670/90, CERN preprint and references therein. 

\item {[11]} B. W. Lee, C. Quigg, H. B. Thacker, {\it Phys. Rev.}, {\bf D16}
(1977) 1519. 

\bye
\vfil\eject

\covertitle{HOW TO PUT A HEAVIER HIGGS\break ON THE LATTICE}
\coverauthor{Urs M. Heller, Herbert~Neuberger, and Pavlos 
Vranas}
\coverpreprint{94}
\coverdate{July 1991}

\makecover
\bye